\documentclass[runningheads]{llncs}
\usepackage[T1]{fontenc}
\usepackage{amsmath}
\usepackage{array}
\usepackage{amssymb}
\usepackage{hyperref}
\usepackage{booktabs}
\usepackage{pifont}
\usepackage{caption}
\usepackage{subcaption}
\usepackage{graphicx}
\usepackage[misc,geometry]{ifsym}

\usepackage{color}

\begin{document}
\title{Unsupervised dMRI Artifact Detection via Angular Resolution Enhancement and Cycle Consistency Learning}
\titlerunning{UdAD-AC}
%
\author{
Sheng Chen\inst{1,2} \and 
Zihao Tang\textsuperscript{1,2,3(\Letter)} \and 
Xinyi Wang\inst{1,2} \and 
Chenyu Wang\inst{2,3,4} \and
Weidong Cai\inst{1}
}

\institute{School of Computer Science, University of Sydney, NSW 2008, Australia \\
\email{zihao.tang@sydney.edu.au}\and
Brain and Mind Centre, University of Sydney, NSW 2050, Australia\and
Central Clinical School, University of Sydney, NSW 2050, Australia\and
School of Biomedical Engineering, University of Sydney, NSW 2008, Australia
}

\authorrunning{S. Chen et al.}
\maketitle             
\begin{abstract}
Diffusion magnetic resonance imaging (dMRI) is a crucial technique in neuroimaging studies, allowing for the non-invasive probing of the underlying structures of brain tissues. Clinical dMRI data is susceptible to various artifacts during acquisition, which can lead to unreliable subsequent analyses. Therefore, dMRI preprocessing is essential for improving image quality, and manual inspection is often required to ensure that the preprocessed data is sufficiently corrected. However, manual inspection requires expertise and is time-consuming, especially with large-scale dMRI datasets. Given these challenges, an automated dMRI artifact detection tool is necessary to increase the productivity and reliability of dMRI data analysis. To this end, we propose a novel unsupervised deep learning framework called \textbf{U}nsupervised \textbf{d}MRI \textbf{A}rtifact \textbf{D}etection via \textbf{A}ngular Resolution Enhancement and \textbf{C}ycle Consistency Learning (UdAD-AC). UdAD-AC leverages dMRI angular resolution enhancement and cycle consistency learning to capture the effective representation of artifact-free dMRI data during training, and it identifies data containing artifacts using designed confidence score during inference. To assess the capability of UdAD-AC, several commonly reported dMRI artifacts, including bias field, susceptibility distortion, and corrupted volume, were added to the testing data. Experimental results demonstrate that UdAD-AC achieves the best performance compared to competitive methods in unsupervised dMRI artifact detection. The code for the the proposed UdAD-AC is available at \url{https://mri-synthesis.github.io/}.

\end{abstract}
\section{Introduction}
Diffusion magnetic resonance imaging (dMRI) is sensitized to Brownian motion, allowing for the assessment of water diffusivity constrained by tissue structures~\cite{mcrobbie2017mri}. dMRI has emerged as a critical tool in neuroimaging studies since the 1990s, providing invaluable insights into brain structures and pathology~\cite{tsuruda1990diffusion}. dMRI data consists of images acquired without any diffusion-sensitizing gradient (usually referred to as $b_{0}$ image) and diffusion-weighted images (DWIs)~\cite{tournier2019diffusion}. Clinical dMRI data is prone to various artifacts that arise from subjects, hardware, software, and environmental factors~\cite{mcrobbie2017mri,tang75reducing,tang2022tw}. Among these artifacts, bias field, susceptibility distortion, and corrupted volume are commonly reported and significantly degrade image quality, necessitating rigorous preprocessing to mitigate their impacts~\cite{tax2022s}. When these artifacts remain undetected or insufficiently restored, the validity of subsequent analyses is compromised, resulting in unreliable measures~\cite{tournier2019mrtrix3}.

Given the significance of artifact-free dMRI data, manual inspection of the preprocessed data is required to ensure its reliability. Manual inspection process is labor-intensive, requires expertise, and is infeasible for large-scale dMRI datasets. Besides, manual inspection is subjective, leading to high inter-rater variability~\cite{victoroff1994method}. Due to aforementioned reasons, an automated artifact detection tool is warranted for dMRI preprocessing pipeline. Such a tool would not only alleviate the defects of manual inspection but also enhance the productivity and reliability of dMRI studies. Existing dMRI quality checking tools have provided statistical-based indices for automated artifact detection, such as signal-to-noise ratio, contrast-to-noise ratio, and normalized correlation between slices~\cite{tournier2019mrtrix3}. These statistical-based indices are predefined to capture only specific artifacts and lack consensus~\cite{tax2022s}. Several deep learning approaches, trained on annotated dMRI artifact data, have shown promise in detecting mixed artifact patterns. QC-Automator~\cite{samani2020qc} employs decoders to identify artifacts present in axial and sagittal 2D slices, improving artifact detection in each respective plane. Extending the QC-Automator, two recent studies~\cite{ettehadi2022automated,ahmad20233d} employ squeeze-and-excitation and dense connections to detect artifacts in 3D volumes, improving the continuity of artifact detection across volumes. However, the reliance on large annotated datasets limits the applicability of such supervised-based methods, as obtaining manual annotations for dMRI artifacts is often impractical.

To address these limitations, we propose a novel unsupervised deep learning framework called \textbf{U}nsupervised \textbf{d}MRI \textbf{A}rtifact \textbf{D}etection via \textbf{A}ngular Resolution Enhancement and \textbf{C}ycle Consistency Learning (UdAD-AC). This framework is designed to detect artifacts in dMRI data without the need for annotated artifacts during training. During the training process, UdAD-AC transforms artifact-free dMRI volumes (a $b_0$ image plus 6 unique DWIs) into an angular resolution enhanced fractional anisotropy (FA) map. This enhanced FA map is then mapped back into averaged input dMRI volumes using cycle consistency learning, ensuring translation consistency at both the image and feature levels. At the inference stage, UdAD-AC, having learned only the representation of artifact-free data, fails the transformation on dMRI volumes containing artifacts. This failure implies the presence of artifacts and can be quantitatively identified by the designed confidence score. To assess the capability of UdAD-AC, commonly reported artifacts such as bias field, susceptibility distortion, and corrupted volume were added into the testing data. Experimental results on the public dataset demonstrate the effectiveness of UdAD-AC, showing it outperforms competitive methods in detecting dMRI artifacts. Our contributions can be summarized as follows:
\begin{itemize}
    \item A novel unsupervised dMRI artifact detection framework, UdAD-AC, is proposed via angular resolution enhancement and cycle consistency learning.
    \item Validated on a public dataset, UdAD-AC outperforms competitive methods across all metrics in dMRI artifact detection.
    \item UdAD-AC increases the reliability and productivity of dMRI studies and has the potential to serve as an automated dMRI quality control tool.
\end{itemize}

\section{Method}
The goal of unsupervised dMRI artifact detection is to train a model to identify dMRI artifacts using only artifact-free dMRI data. Such data is available in many public dMRI datasets that have been preprocessed with proper quality checking. During training, the proposed UdAD-AC employs angular resolution enhanced FA map to capture the representation of artifact-free dMRI data, and then utilizes cycle consistency learning to enhance the compactness of learned representation. During testing, UdAD-AC identifies dMRI data containing artifacts using a thresholding confidence score.
 \begin{figure}
    \centering     
    \includegraphics[width=\textwidth]{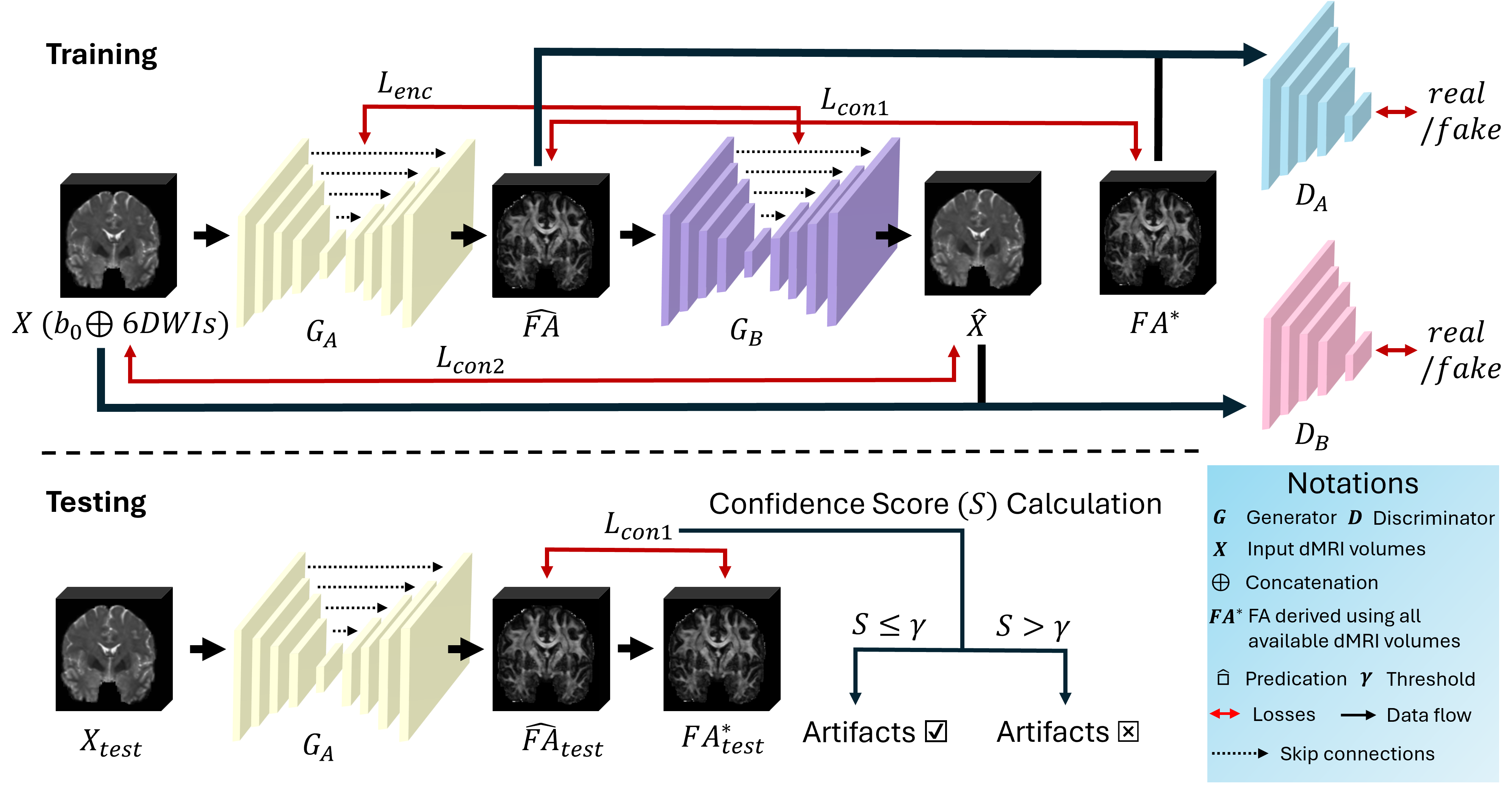}
    \caption{The detailed framework of the proposed UdAD-AC.}
    \label{fig:net}
\end{figure}

The detailed framework of UdAD-AC is illustrated in Figure~\ref{fig:net}. UdAD-AC consists of generators $G_{A}$ and $G_{B}$, along with corresponding discriminators $D_{A}$ and $D_{B}$.  $G_{A}$ takes a $b_0$ image plus 6 unique DWIs as input volumes ($X$) to generate an angular resolution enhanced FA map ($\widehat{FA}$). This enhanced FA map is then fed into $G_{B}$ to produce averaged input dMRI volumes ($\widehat{X}$). The discriminators $D_A$ and $D_B$ distinguish $\widehat{FA}$ and $\widehat{X}$ from their real counterparts, $FA^*$ and $X$, respectively.

\subsection{Angular Resolution Enhancement}
The dMRI data of a subject typically includes a large number of volumes, comprising a $b_0$ image and unique DWIs acquired from diffusion-sensitizing gradient directions. In neuroimaging studies, dMRI data can be analyzed using the diffusion tensor imaging (DTI) model to characterize brain structures~\cite{tournier2011diffusion}. The DTI model is denoted as a symmetrical matrix $\mathcal{D}$ and the coefficients in $\mathcal{D}$ are estimated from dMRI data by solving the Stejskal-Tanner Equation as:
\begin{align}
    \label{eq:dti}
    S_{k}=S_{0}e^{-b \hat{g}^T_k \mathcal{D} \hat{g}_{k}}, & & \medskip \mathcal{D} = 
    \begin{bmatrix}
    \mathcal{D}_{xx} & \mathcal{D}_{xy} & \mathcal{D}_{xz} \\
    \mathcal{D}_{xy} & \mathcal{D}_{yy} & \mathcal{D}_{yz} \\
    \mathcal{D}_{xz} & \mathcal{D}_{yz} & \mathcal{D}_{zz} \\
    \end{bmatrix},
\end{align}
where $S_0$ is the signal usually referred to as the $b_{0}$ image and $S_k$ is the diffusion weighted signal (usually referred to as the DWI) acquired along the gradient direction $\hat{g}_{k}$ with diffusion sensitization value $b$~\cite{o2011introduction,tournier2011diffusion}.
A diffusion metric, FA, can be further derived from $\mathcal{D}$ to quantitatively summarize the diffusivity within brain voxels~\cite{o2011introduction}. Compared to other diffusion metrics, FA is a ratio of variances normalized by the total diffusion, which is a dimensionless measure naturally within the scale of 0 to 1~\cite{o2011introduction}. The formulation of FA is denoted as follows:
\begin{equation}
    \label{eq:fa}
    \text{FA} =  \sqrt{\frac{(\lambda_1-\lambda_2)^2 + (\lambda_1-\lambda_3)^2 + (\lambda_2-\lambda_3)^2}{2*(\lambda_1^2 + \lambda_2^2 + \lambda_3^2)}},
\end{equation}
where $\lambda_1$, $\lambda_1$, and $\lambda_3$ are eigenvalues of $\mathcal{D}$.  

\begin{figure}
     \centering
     \begin{subfigure}[b]{0.3\textwidth}
         \centering
         \includegraphics[width=4cm,height=3cm]{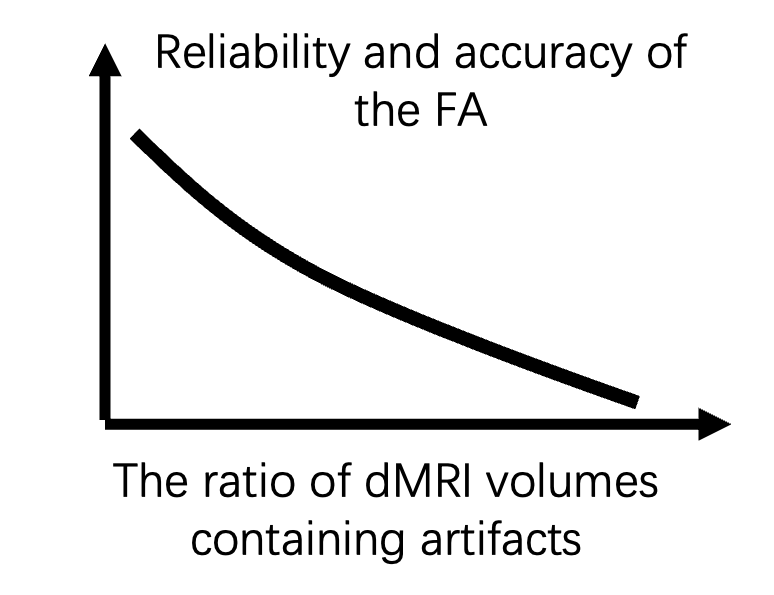}
         \caption{\label{fig:distributionA}}
         \label{fig:distA}
     \end{subfigure}
     \hfill
     \begin{subfigure}[b]{0.3\textwidth}
         \centering
         \includegraphics[width=4cm,height=3cm]{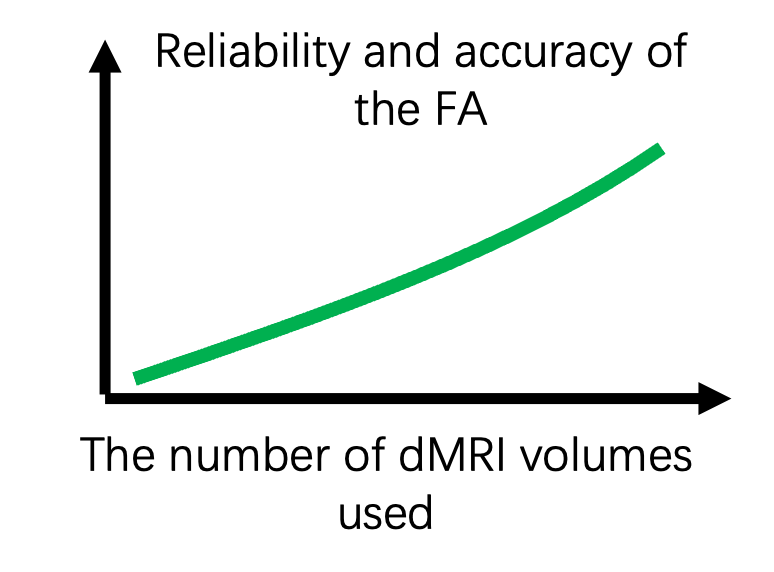}
         \caption{\label{fig:distributionB}}
         \label{fig:distB}
     \end{subfigure}
     \hfill
     \begin{subfigure}[b]{0.3\textwidth}
         \centering
         \includegraphics[width=4cm,height=3cm]{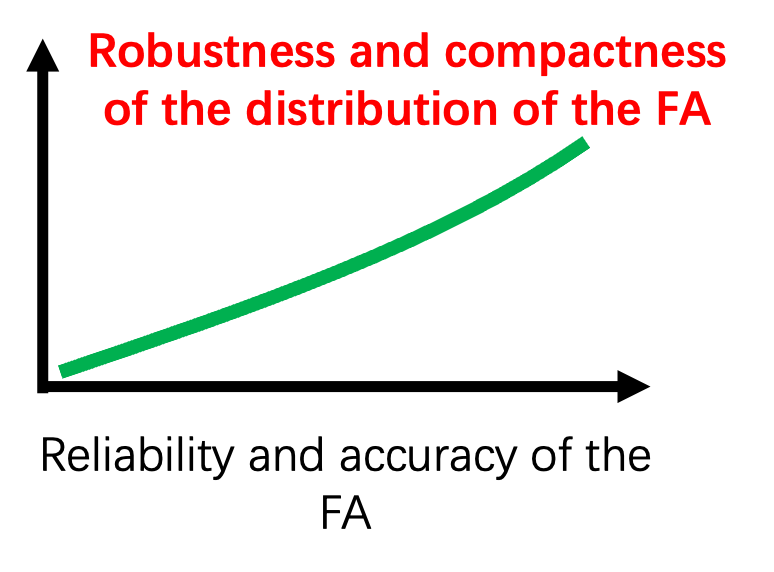}
         \caption{\label{fig:distributionC}}
         \label{fig:distC}
     \end{subfigure}
        \caption{An illustration of relationships between the derived FA and the dMRI volumes used for estimation.}
        \label{fig:distribution}
\end{figure}

The presence of artifacts in dMRI volumes alters the diffusion information, leading to differences in the derived FA map compared to that obtained from artifact-free dMRI volumes. However, since artifacts may only be present in a single volume within a large set of otherwise artifact-free volumes~\cite{tax2022s}, their impact on the overall diffusion information can be minimal. This subtle change makes it difficult to detect the artifacts by simply analyzing the FA map distribution. Using a smaller set of dMRI volumes to derive the corresponding FA map can address this issue, as the relative impact of any single volume with artifacts becomes more pronounced when focusing on a smaller subset. Theoretically, a minimum set of a $b_0$ image and 6 unique DWIs is required to estimate DTI and then derive FA~\cite{o2011introduction}. However, the reliability and accuracy of the derived FA also depend on the number of unique dMRI volumes used, known as angular resolution~\cite{ni2006effects}. The higher the angular resolution (i.e., the more unique dMRI volumes used), the more reliable the resulting FA distribution will be. Figure~\ref{fig:distribution} is provided to help illustrate relationships between these concepts.

Considering all the above factors, the task of dMRI artifact detection has been formulated as angular resolution enhancement. This approach utilizes limited dMRI volumes to generate reliable diffusion metrics comparable to those derived from sufficient dMRI volumes. The feasibility of angular resolution enhancement has been explored in recent studies~\cite{tang2023high,aja2023validation,chen2024enhancing}. In the training process, the generator $G_{A}$ accepts artifact-free $b_0$ image concatenated with 6 unique DWIs ($b_0 \oplus 6DWIs$) of a subject as inputs $X\in\mathbb{R}^{N\times 7\times W \times H \times D}$ and transforms $X$ into an angular resolution enhanced FA map, denoted as $\widehat{FA}\in \mathbb{R}^{N\times 1\times W \times H \times D}$. It is supervised by the $FA^*$ (the FA map derived from all available dMRI volumes of the subject), using the loss $L_{con1}$. The objective function $L_{con1}$ can be formulated as Equation~\ref{eq:lcon1}.
\begin{equation}
    \label{eq:lcon1}
    L_{con1} = ||\widehat{FA} - FA^*||_{1} = ||\phi(G_{A}(X)) - FA^*||_{1}, \\ 
\end{equation} 
where $\phi$ denotes the Sigmoid function. Using only artifact-free dMRI volumes, the designed angular resolution enhancement facilitates $G_A$ to capture the distribution of $FA^*$, whose representation is naturally compact and robust. During the testing, when the enhanced FA is significantly different from the learned representation, the model identifies the presence of the artifacts.  

\subsection{Cycle Consistency Learning}
To ensure the model effectively exploits the diffusion information from $X$ and further constrain the space of possible mappings from $X$ to $\widehat{FA}$, UdAD-AC introduces cycle consistency learning (CCL)~\cite{zhu2017unpaired,zhao2021anomaly} to ensure translation consistency at both the image and feature levels. 

The generator $G_A$ transforms $X$ into the $\widehat{FA}$ and the generator $G_B$ maps $\widehat{FA}$ into $\widehat{X}\in\mathbb{R}^{N\times 2\times W \times H \times D}$, comprising a predicted $\widehat{b_0}$ image and a synthesized $\widehat{DWI_\xi}\in\mathbb{R}^{N\times 1\times W \times H \times D}$ volume containing the averaged diffusion information from 6 input DWIs ($6DWIs$). The $\widehat{X}$ is supervised by inputs $X$ using the loss $L_{con2}$ as follows:
\begin{equation}
    L_{con2} = ||\widehat{b_0} - b_0||_2 + ||\widehat{DWI_\xi} - \xi( 6DWIs)||_2,
\end{equation}
where $\xi$ denotes the averaging operation along the volume. The $b_0$ image primarily provides spatial information, and the averaged diffusion information from 6 input DWI volumes serves as a central constraint~\cite{wen2016discriminative}, encouraging the model to extract a compact representation of the input diffusion information. Additionally, inspired by~\cite{zhao2021anomaly,akcay2019ganomaly}, UdAD-AC constrains the feature translation consistency between $E_A$ and $E_B$, which are encoders of $G_A$ and $G_B$ respectively. The feature translation consistency is achieved by the loss $L_{enc}$:
\begin{equation}
    L_{enc} = \frac{\sum^n_{i=0} || E_A(X)_{i} -  E_B(\widehat{FA})_{i}||_2}{n},
\end{equation}
where $i$ denotes the layer position and $n$ is the number of layers. Similar to Cycle-GAN~\cite{zhu2017unpaired}, discriminators $D_A$ and $D_B$ are employed in the training to discriminate the generated $\widehat{FA}$ and $\widehat{X}$ from $FA^*$ and $X$, respectively. The adversarial learning facilitates generators to produce images that approach a similar distribution as real ones, which can be formulated using least squares as follows:
\begin{equation}
    \label{eq:ladv}
    L_{adv} = L_{D_A} + L_{D_B},
\end{equation}
\begin{equation}
    L_{D_A} = ||D_A(FA^*)) - 1||_2 + ||D_{A}(\widehat{FA})||_2,
\end{equation}
\begin{equation}
    L_{D_B} = ||D_B(b_0 \oplus \xi(6DWIs)) - 1||_2 + ||D_{B}(\widehat{X})||_2,
\end{equation}
where $\oplus$ denotes concatenation, $\xi$ denotes the averaging operation along the volume. By exploring CCL, UdAD-AC learns a translation consistent representation of $X$ and $FA^*$, which compactly and effectively represents the distribution of artifact-free dMRI data. Consequently, CCL facilitates the model's ability to identify artifacts during testing, as dMRI artifacts exhibit significantly distinct representations at both the image and feature levels. In summary, UdAD-AC is trained to minimize the generative loss terms $L_{gen}$ as shown in Equation~\ref{eq:lgen} and to maximize the adversarial loss $L_{adv}$:
\begin{equation}
\label{eq:lgen}
    L_{gen} = \alpha_{1}L_{con1} + \alpha_{2}L_{con2} +  \alpha_{3}L_{enc},
\end{equation}
where $\alpha_1$, $\alpha_2$, and $\alpha_3$ are the loss weights.

\subsection{Model Testing}
During the testing stage, the model uses $L_{con1}$ in Equation~\ref{eq:lcon1} to calculate the confidence score of a given image. Given a test sample $X_{test}$ and the $FA^*_{test}$, the confidence score ($S$) is defined as:
\begin{equation}
\label{eq:S}
    S = 1- \phi(\frac{||\phi(G_A(X_{test})) - FA^*_{test}||_1 - \mu_{train}}{\sigma_{train}}),
\end{equation}
where $\mu_{train}$ and $\sigma_{train}$ are the mean and standard deviation of $L_{con1}$ calculated on training samples, $\phi$ is the Sigmoid function to ensure $S$ is within the scale of 0 to 1. When $S < \gamma$, the test sample is identified as the dMRI data containing artifacts. 

\section{Experiments and Results}

\subsection{Dataset and Data Preparation}
The public Human Connectome Project (HCP) database~\cite{van2013wu} is adopted for this study. Following~\cite{tang2023high}, the extracted HCP database comprises single-shell, preprocessed artifact-free dMRI data with proper quality checking. Each subject contains a $b_0$ image and 90 unique DWIs acquired with a diffusion sensitization value of $b=1000$. A total of 100 subjects are randomly selected from the extracted HCP data and split into training (50 subjects) and testing (50 subjects) sets. Among the 50 testing subjects, 20 subjects are randomly selected to add synthesized dMRI artifacts, including bias field, susceptibility distortion, and corrupted volume. The bias field is created by multiplying dMRI volumes with localized Gaussian noise. The susceptibility distortion is synthesized using elastic transformation. Corrupted volumes are generated by replacing random dMRI volumes with zeros. These three highlighted artifacts are commonly reported and significantly degrade image quality if left undetected or insufficiently restored~\cite{tax2022s}. All available dMRI volumes of each subject are used to derive the corresponding FA map with the highest possible angular resolution, denoted as $FA^*$. For each subject, a $b_0$ image plus 6 unique DWIs are sub-sampled from the 90 unique DWIs to serve as inputs for the network.

\subsection{Implementation Details}
Both generators of UdAD-AC utilize a vanilla encoder-decoder architecture, and skip connections are adopted to transmit intermediate encoder features to the decoder. The discriminators share the same architecture as the encoder. The Adam optimizer is employed with an initial learning rate of $1 \times 10^{-3}$, and all compared deep learning models are trained for 100 epochs. Input images are resized to $160 \times 192 \times 160$ and fed to the UdAD-AC with a batch size of 4. UdAD-AC is optimized based on the weighted loss $L_{gen}$ using the weight values $\alpha_1 = 50$, $\alpha_2 = 10$, and $\alpha_3 = 1$ in practice~\cite{akcay2019ganomaly}. The threshold $\gamma$ for the confidence score in Equation~\ref{eq:S} is set to 0.1. All experiments were run on a single NVIDIA GeForce Tesla V100-SXM2 GPU with 32GB of memory, and the models were implemented with PyTorch 2.0.0 and Python 3.10.12.

\subsection{Experimental Results}
To quantitatively evaluate the performance of the proposed UdAD-AC and competitive methods, the accuracy (ACC), F1 score (F1), sensitivity (SEN), specificity (SPE), and area under the curve (AUC) are summarized in Table~\ref{table:main}.
\begin{table}[!htb]
\centering
\begin{tabular}{c|c|c|c|c|c}
\hline \hline
 & \multicolumn{5}{c}{\textbf{HCP Testing}} \\ \hline \hline
Method & ACC (\%) & F1 (\%) & SEN (\%)  & SPE  (\%) & AUC (\%)\\ \hline
VAE~\cite{kingma2013auto} & 80 & 84.84 & 93.33 & 69 & 76.67 \\ \hline
f-AnoGAN~\cite{schlegl2019f} & 84 & 87.50 & 93.33 & 70 & 81.67 \\  \hline
Ganomaly~\cite{akcay2019ganomaly} & 86 & 89.23 & 96.67 & 75 & 84.17 \\  \hline
SALAD~\cite{zhao2021anomaly} & 86 & 88.89  & 93.33 & 70 & 83.33 \\ \hline \hline
UdAD-AC (w/o CCL) & 92 & 93.54 & 96.67 & 85 & 90.83 \\  \hline
UdAD-AC & \textbf{96} & \textbf{96.67} & \textbf{96.67} & \textbf{95} & \textbf{95.83} \\   \hline
\end{tabular}
\caption{Quantitative comparisons between the proposed UdAD-AC and competitive methods on the HCP testing set. The accuracy (ACC), F1 score (F1), sensitivity (SEN), specificity (SPE), and area under the curve (AUC) are reported. The w/o CCL denotes the model without the cycle consistency learning.}
\label{table:main}
\end{table}
From the results, it can be observed that the proposed method achieves the best performance across all metrics. The competitive unsupervised artifact detection methods~\cite{kingma2013auto,schlegl2019f,akcay2019ganomaly,zhao2021anomaly} commonly employ auto-encoding, which encodes inputs into a compressed representation and then decodes it back to the reconstructed inputs~\cite{michelucci2022introduction}. While these methods work well in the context of 2D modalities, they encounter issues when adapted to 3D dMRI artifact detection. Given input 3D dMRI volumes with complicated brain anatomical structures, the spatial information of dMRI volumes is lost in the deep layers of the encoder~\cite{he2016deep}, making input reconstruction infeasible. Thus, skip connections~\cite{ronneberger2015u} are necessarily used to transmit intermediate encoder features to the decoder. However, the adoption of skip connections inevitably results in the network learning an identity mapping function, which hinders its capacity in detecting dMRI artifacts that alter anatomical structures, such as susceptibility distortion. Confidence scores calculated by the compared algorithms for a set of input dMRI volumes of a subject with susceptibility distortion are shown in Table~\ref{table:sus}.
\begin{table}[!htb]
    \centering
    \begin{tabular}{c|c|c|c|c|c} \hline \hline
                &  \multicolumn{5}{c}{Susceptibility Distortion}\\ \hline \hline
    Method         &  VAE & f-AnoGAN & Ganomaly & SALAD & UdAD-AC \\ \hline
    Confidence  &  0.51 &  0.46    & 0.39     & 0.33 & 0.04 \\ \hline
    Classification & \ding{55}  & \ding{55} & \ding{55} & \ding{55} & \ding{51} \\ \hline
    \end{tabular}
    \caption{Confidence scores calculated by compared methods for a subject chosen from the HCP testing set, which has susceptibility distortion. When the confidence score $<0.1$, the current input dMRI volumes are identified as containing artifacts. \ding{51} denotes the correct classification.}
    \label{table:sus}
\end{table}
Instead of auto-encoding, the proposed UdAD-AC formulates the unsupervised dMRI artifact detection task as angular resolution enhancement, which avoids identity mapping and efficiently captures the distribution of artifact-free dMRI data. Furthermore, the designed cycle consistency learning enhances the robustness and compactness of the learned representation. 

\begin{figure}[!htb]
    \begin{minipage}[b]{0.16\textwidth}
        \centering
        \centerline{\includegraphics[width=2cm,height=2.4cm]{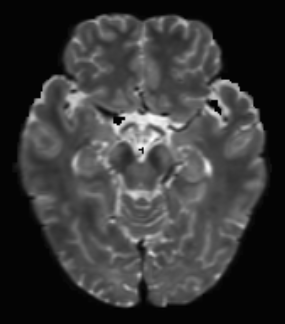}}
        \centerline{(a) $X_1$}
    \end{minipage}
    \hfill
    \begin{minipage}[b]{0.16\textwidth}
        \centering
        \centerline{\includegraphics[width=2cm,height=2.4cm]{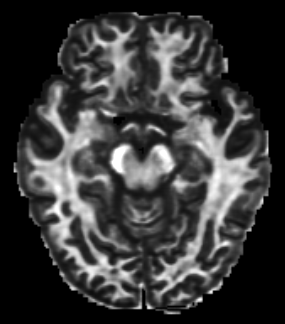}}
        \centerline{(b) $\widehat{FA}_{X_1}$}
    \end{minipage}
    \hfill
    \begin{minipage}[b]{0.16\textwidth}
        \centering
        \centerline{\includegraphics[width=2cm,height=2.4cm]{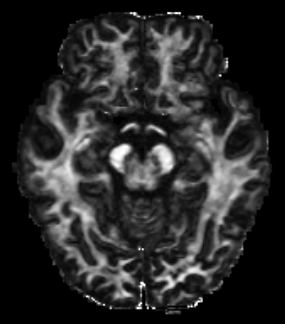}}
        \centerline{(c) $FA^*_{X_1}$}
    \end{minipage}
    \hfill
    \begin{minipage}[b]{0.16\textwidth}
        \centering
        \centerline{\includegraphics[width=2cm,height=2.4cm]{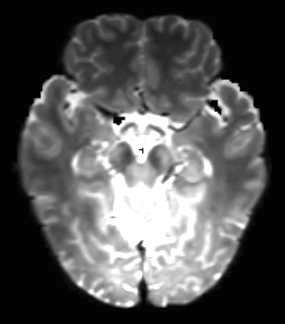}}
        \centerline{(d) $X_2$}
    \end{minipage}
    \hfill
    \begin{minipage}[b]{0.16\textwidth}
        \centering
        \centerline{\includegraphics[width=2cm,height=2.4cm]{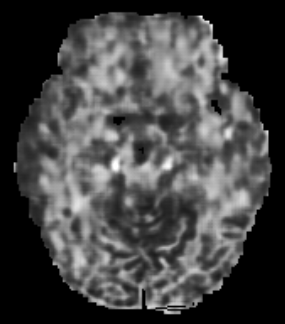}}
        \centerline{(e) $\widehat{FA}_{X_2}$}
    \end{minipage}
    \hfill
    \begin{minipage}[b]{0.16\textwidth}
        \centering
        \centerline{\includegraphics[width=2cm,height=2.4cm]{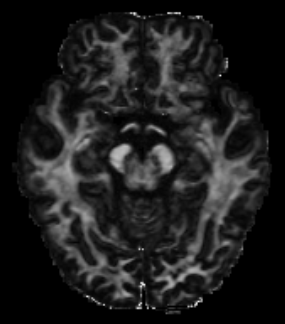}}
        \centerline{(f) $FA^*_{X_2}$}
    \end{minipage}
    \caption{Qualitative comparisons in axial view of two sets of dMRI data and corresponding predicted FA maps by UdAD-AC. $\widehat{FA}_{X_1}$ is generated from artifact-free $X_1$, $FA^*_{X_1}$ is derived from all artifact-free dMRI volumes (including $X_1$) of the subject, $\widehat{FA}_{X_2}$ is generated from $X_2$, and $FA^*_{X_2}$ is derived from all available dMRI volumes (including $X_2$) of the subject.}
    \label{fig:bias}
\end{figure}

To demonstrate the efficiency of UdAD-AC in amplifying the discrepancy between representations of artifact-free dMRI data and dMRI data containing artifacts, two identical sets of input dMRI volumes of a subject are chosen from the testing set and fed to UdAD-AC with different operations. The first set $X_1$, remains artifact-free, and the second set $X_2$, has an applied bias field. The qualitative comparisons of the two sets of dMRI data and the corresponding predicted FA maps by UdAD-AC are demonstrated in Figure~\ref{fig:bias}
, where $\widehat{FA}_{X_1}$ is generated from $X_1$, $FA^*_{X_1}$ is derived by using all artifact-free dMRI volumes (including $X_1$) of the subject, $\widehat{FA}_{X_2}$ is generated from $X_2$, and $FA^{*}_{X_2}$ is derived by using all available dMRI volumes (including $X_2$) of the subject. From Figure~\ref{fig:bias}, it can be observed that $\widehat{FA}_{X_2}$ shows a rough appearance and is significantly different from $\widehat{FA}_{X_1}$. $\widehat{FA}_{X_1}$ is smooth and shares the similar appearance to $FA^*_{X_1}$ and $FA^*_{X_2}$. This significant difference is rooted in the distinct representation of $X_1$ and $X_2$ and is amplified by the proposed angular resolution enhancement with cycle consistency, helping the model easily identify the presence of artifacts. 

\subsubsection{Detection of Corrupted Volumes}
\begin{figure}[!htb]
    \centering
    \begin{minipage}[b]{0.32\textwidth}
        \centering
        \centerline{\includegraphics[width=4cm,height=4.8cm]{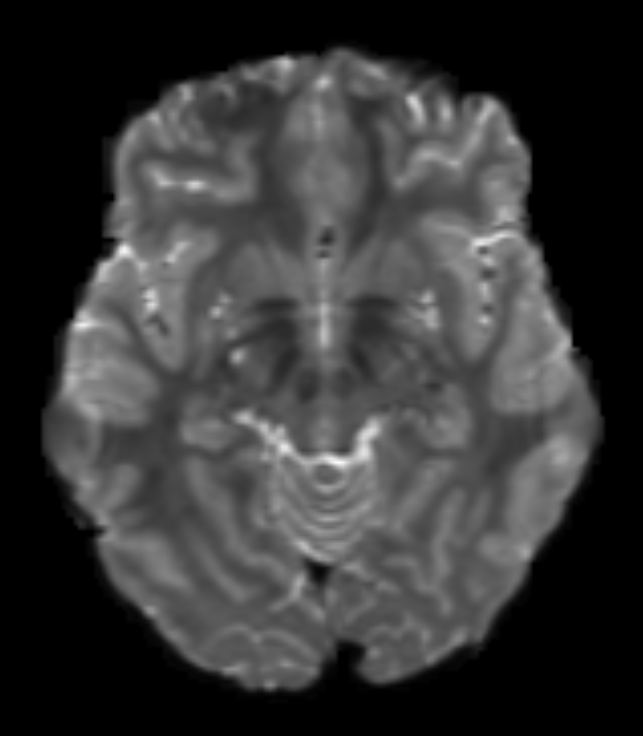}}
        \centerline{(a) $X_{corr}$}
    \end{minipage}
    \hfill
    \begin{minipage}[b]{0.32\textwidth}
        \centering
        \centerline{\includegraphics[width=4cm,height=4.8cm]{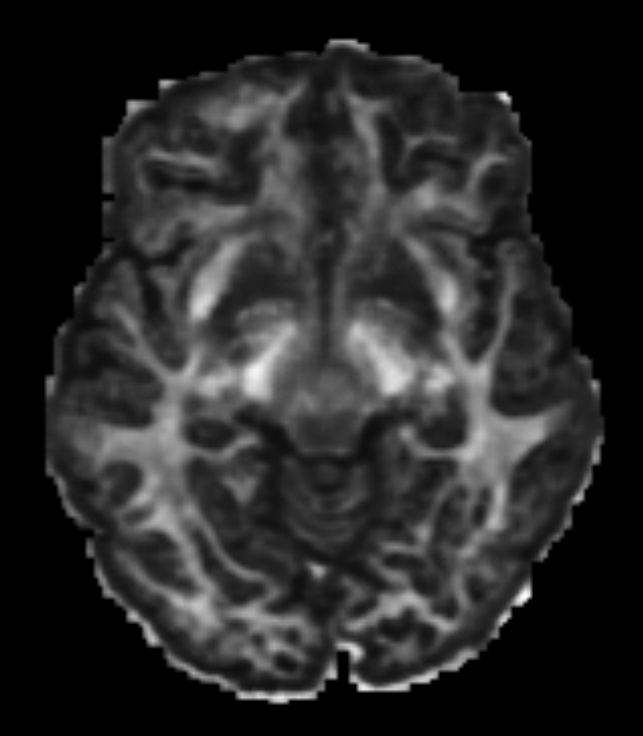}}
        \centerline{(b) $\widehat{FA}_{corr}$}
    \end{minipage}
    \hfill
    \begin{minipage}[b]{0.32\textwidth}
        \centering
        \centerline{\includegraphics[width=4cm,height=4.8cm]{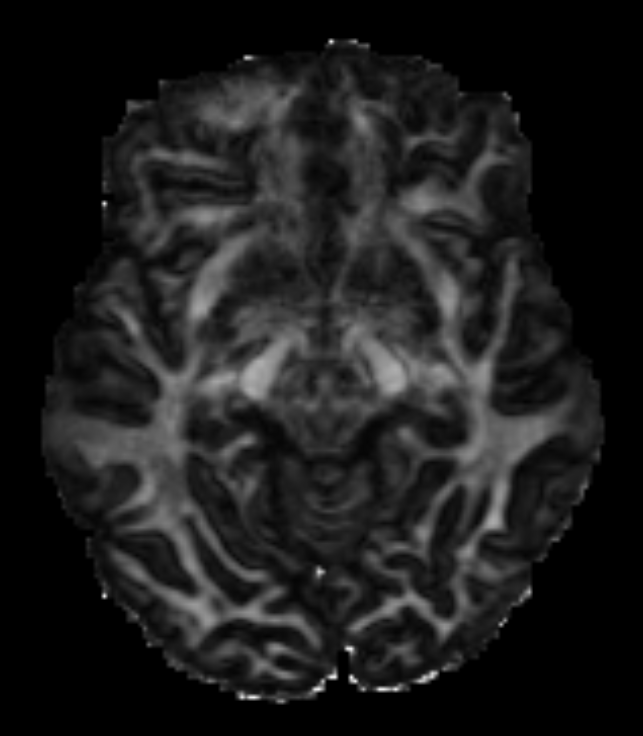}}
        \centerline{(c) $FA^*_{corr}$}
    \end{minipage} 
    \caption{Visual comparison of FA maps generated from dMRI data with corrupted volumes. (a) $X_{corr}$: dMRI data with corrupted volumes. (b) $\widehat{FA}_{corr}$: predicted FA map generated from the corrupted dMRI data, showing a blurry appearance. (c) $FA^*_{corr}$: FA map derived using all available dMRI volumes, showing clear anatomical details.}
    \label{fig:corrupted}
\end{figure}

 Corrupted volumes are expected to be captured by UdAD-AC regardless of their position or number. Failure to do so would significantly reduce the capability of UdAD-AC. To validate that UdAD-AC can robustly capture the presence of corrupted volumes, a subject was randomly selected from the HCP testing set. Initially, the first dMRI volume (Volume 0) of the subject was corrupted and then fed to UdAD-AC to produce a confidence score. This process was repeated for each of the remaining dMRI volumes (Volumes 1 to 6). The results are shown in Table~\ref{table:corv}, demonstrating that UdAD-AC can identify the presence of corrupted volumes whenever they persist in the input dMRI data. Additionally, Figure~\ref{fig:corrupted} provides a qualitative comparison between the FA maps generated from dMRI data with corrupted volumes ($X_{corr}$). The predicted FA map from the corrupted data ($\widehat{FA}_{corr}$) appears blurry, while the FA map derived by using all available dMRI volumes ($FA^*_{corr}$) retains clear anatomical details, further demonstrating UdAD-AC's robustness in detecting corrupted volumes.
\begin{table}[!htb]
    \centering
    \begin{tabular}{c|c|c|c|c|c|c|c} \hline \hline
         &  \multicolumn{7}{c}{Corrupted Volume} \\ \hline \hline
    Position & Volume 0 & Volume 1 & Volume 2 & Volume 3 & Volume 4 & Volume 5 & Volume 6 \\ \hline
    Confidence & 0.03 & 0.05 & 0.05 & 0.05 & 0.07 & 0.03 & 0.06 \\ \hline
    Classification & \ding{51} & \ding{51} & \ding{51} & \ding{51} & \ding{51} & \ding{51} & \ding{51} \\ \hline
    \end{tabular}
    \caption{Confidence scores produced by UdAD-AC for a subject with iteratively corrupted dMRI volumes. When the confidence score $<0.1$, the corrupted volume is identified. \ding{51} denotes the correct classification.}
    \label{table:corv}
\end{table}

\subsubsection{Detection of Susceptibility Distortion}
\begin{figure}
    \centering
    \begin{minipage}[b]{0.32\textwidth}
        \centering
        \centerline{\includegraphics[width=3.8cm,height=3.8cm]{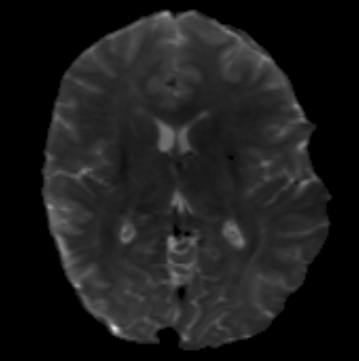}}
        \centerline{(a) $X_{dist}$}
    \end{minipage}
    \begin{minipage}[b]{0.32\textwidth}
        \centering
        \centerline{\includegraphics[width=3.8cm,height=3.8cm]{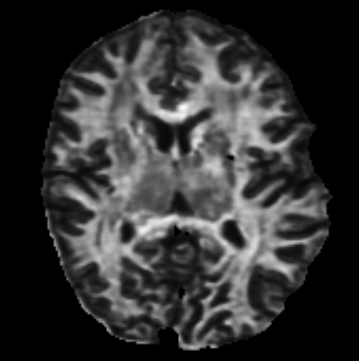}}
        \centerline{(b) $\widehat{FA}_{dist}$}
    \end{minipage}
    \begin{minipage}[b]{0.32\textwidth}
        \centering
        \centerline{\includegraphics[width=3.8cm,height=3.8cm]{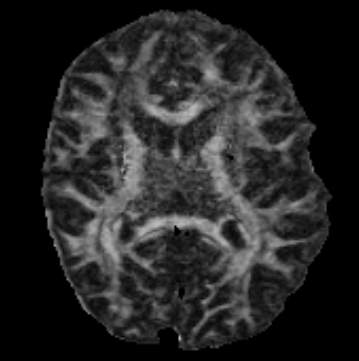}}
        \centerline{(c) $FA^*_{dist}$}
    \end{minipage} 
    \caption{Qualitative examples of a subject with susceptibility distortion in axial view . a) $X_{dist}$ is the input dMRI data to the UdAD-AC, b) $\widehat{FA}_{dist}$ is the predicted FA generated from $X_{dist}$, and c) $FA^*_{dist}$ is the FA derived by using all dMRI volumes of the subject.}
    \label{fig:sus}
\end{figure}

Unlike artifacts that only occur in partial volumes, susceptibility distortion ($dist$) impacts all available dMRI volumes of the subject. As a result, the diffusion information in ${FA^*_{dist}}$, which is derived using all available dMRI volumes of the subject with susceptibility distortion, is significantly affected. In this situation, UdAD-AC may struggle to identify the presence of susceptibility distortion by comparing the reconstruction error between $\widehat{FA}_{dist}$ (generated by UdAD-AC using distorted dMRI data $X_{dist}$) and $FA^*_{dist}$. This is because both $\widehat{FA}_{dist}$ and $FA^*_{dist}$ are significantly alerted by susceptibility distortion and may have similar representations at both the image and feature levels. However, experimental results show that UdAD-AC still accurately identifies the presence of susceptibility distortion through a significant reconstruction error between $\widehat{FA}_{dist}$ and $FA^*_{dist}$. This indicates $\widehat{FA}_{dist}$ and $FA^*_{dist}$ have different representations at both image and feature levels.
A qualitative example of a subject with susceptibility distortion is presented in Figure~\ref{fig:sus}.

\subsection{Ablation Studies}
\subsubsection{Translation Consistency}
The effectiveness of CCL is demonstrated in Table~\ref{table:main}. CCL facilitates UdAD-AC in learning translation consistent representations and effectively utilizing the diffusion information from artifact-free input dMRI volumes ($X$). During the design of CCL, the outputs of generator $G_B$ were explored to find the most effective configuration. The performance of these configurations is shown in Table~\ref{table:abl}, where $\widehat{b_0}$ is the predicted $b_0$ image, $\widehat{6DWIs}$ are the predicted 6 input DWIs, $\widehat{DWI_\xi}$ is the synthesized DWI containing averaged diffusion information from $6DWIs$, and $\oplus$ denotes concatenation. The table shows that $\widehat{b_0}\oplus\widehat{DWI_\xi}$ achieves the best performance. We consider this due to the following reason. Since $\widehat{FA}$ has higher angular resolution diffusion information than $X$, it can be mapped into various sets of input dMRI volumes. Thus, constraining the translation consistency using $\widehat{6DWIs}$ is not optimal. By averaging diffusion information from $6DWIs$ into $\widehat{DWI_\xi}$, the diffusion information is represented as a centralized representation, which is easier for a model to learn and capture. Therefor, $\widehat{b_0}\oplus\widehat{DWI_\xi}$ is adopted as the output ($\widehat{X}$) of $G_B$.

\subsubsection{Activation Function}
The FA derived from DTI is a ratio of variances normalized by the total diffusion that naturally falls within the scale of 0 to 1~\cite{o2011introduction}. To efficiently explore this property, different activation layers, such as ReLU~\cite{agarap2018deep} and Sigmoid ($\phi$)~\cite{narayan1997generalized}, have been investigated in the calculation of $L_{con1}$ in Equation~\ref{eq:lcon1}. The corresponding performance is summarized in Table~\ref{table:abl}. From the table, it can be observed that $\phi$ achieves the best performance. We assume that $\phi$ facilitates the convergence of the model and efficiently exploits the property of FA. Therefore, $L_{con1}$ adopts $\phi$ as the activation function.

\begin{table}[!htb]
    \centering
    \begin{tabular}{c|c|c|c|c|c} \hline\hline
    \multicolumn{6}{c}{ Outputs of $G_B$ in CCL } \\ \hline\hline
     Combination  & ACC (\%) & F1 (\%) & SEN (\%)  & SPE  (\%) & AUC (\%)\\ \hline
     $\widehat{b_0}\oplus\widehat{6DWIs}$ & 94  & 95.08  & 96.67  & 90  & 93.33 \\ \hline
     $\widehat{b_0}\oplus\widehat{DWI_\xi}$ & \textbf{96} & \textbf{96.67} & \textbf{96.67} & \textbf{95} & \textbf{95.83} \\ \hline  \hline
    \multicolumn{6}{c}{$L_{con1}$ of UdAD-AC } \\ \hline\hline
    Activation & ACC (\%) & F1 (\%) & SEN (\%)  & SPE  (\%) & AUC (\%)\\ \hline
     $-$   & 94  & 95.08  & 96.67  & 90  & 93.33  \\   \hline         
     ReLU  & 94  & 95.08  & 96.67  & 90  & 93.33  \\   \hline
     $\phi$ & \textbf{96} & \textbf{96.67} & \textbf{96.67} & \textbf{95} & \textbf{95.83} \\   \hline
    \end{tabular}
    \caption{Performance comparison of different outputs of $G_B$ in the designed CCL, and different activation functions used in the calculation of $L_{con1}$ for the proposed UdAD-AC. $-$: no activation used; $\phi$: the Sigmoid function.}
    \label{table:abl}
\end{table}

\subsubsection{Weights Balance in $L_{gen}$}
$L_{gen}$ consists of three sub-losses: $L_{con1}$, $L_{con2}$, and $L_{enc}$. Since the core idea of UdAD-AC is to use angular resolution enhancement to capture artifact-free dMRI representations, $L_{con1}$ is given the highest weight intuitively. While $L_{con2}$, the cycle consistency loss, is important for exploiting input dMRI volumes, equal weighting with $L_{con1}$ risks the model overfocusing on input reconstruction, leading to identity mapping. $L_{enc}$ complements by ensuring consistency in the feature space, however, a high coefficient for $L_{enc}$ may overly constrain the intermediate features, resulting in unnatural outputs. Based on these considerations, and following the approach in~\cite{akcay2019ganomaly}, we apply weights of 50, 10, and 1 to $L_{con1}$, $L_{con2}$, and $L_{enc}$, respectively.

\section{Conclusions}
This paper proposes a novel unsupervised deep learning framework, UdAD-AC, which leverages angular resolution enhancement and cycle consistency learning to detect dMRI artifacts without the need for annotated training data. During training, UdAD-AC employs angular resolution enhanced FA map to effectively capture the representation of artifact-free dMRI data. The designed cycle consistency learning enhances the compactness and robustness of the learned representation. During testing, UdAD-AC, quantitatively identifies dMRI data containing artifacts using the designed confidence score. Experiments and ablation studies demonstrate the effectiveness of our approach. UdAD-AC has the potential to serve as part of an automatic dMRI quality control tool, improving the productivity and reliability of dMRI studies and significantly reducing human labor.

\section{Limitations and Future Work}
While UdAD-AC has demonstrated promising results in improving dMRI reliability and reducing manual intervention, several limitations remain. First, this study focuses on three common dMRI artifacts—bias fields, susceptibility distortions, and corrupted volumes. Future work should expand the evaluation to a broader range of artifacts and datasets to enhance the model's generalization. Additionally, the reliance on synthetic artifacts, though practical, may not fully capture the complexity of real-world dMRI data. Future research should focus on validating UdAD-AC on real artifact-affected datasets to ensure its robustness in clinical settings. 

Furthermore, this paper lacks a thorough comparison with the most recent and relevant baseline methods due to the current scope. Future efforts will incorporate more comprehensive comparisons to provide a deeper understanding of UdAD-AC’s performance relative to state-of-the-art techniques.

\subsubsection{Acknowledgments}
The authors acknowledge the support of The BISA Flagship Research Program (2019) and The University of Sydney Office of Global and Research Engagement Catalyst Grants (2024).

\bibliographystyle{splncs04} 
\bibliography{67.bib}

\begin{thebibliography}{10}
\providecommand{\url}[1]{\texttt{#1}}
\providecommand{\urlprefix}{URL }
\providecommand{\doi}[1]{https://doi.org/#1}

\bibitem{agarap2018deep}
Agarap, A.F.: Deep learning using rectified linear units. ArXiv preprint ArXiv:1803.08375  (2018)

\bibitem{ahmad20233d}
Ahmad, A., Parker, D., Dheer, S., et~al.: {3D-QCNet}--{A} pipeline for automated artifact detection in diffusion {MRI} images. Computerized Medical Imaging and Graphics  \textbf{103},  102151 (2023)

\bibitem{aja2023validation}
Aja-Fern{\'a}ndez, S., Mart{\'\i}n-Mart{\'\i}n, C., Planchuelo-G{\'o}mez, {\'A}., et~al.: Validation of deep learning techniques for quality augmentation in diffusion mri for clinical studies. NeuroImage: Clinical p. 103483 (2023)

\bibitem{akcay2019ganomaly}
Akcay, S., Atapour-Abarghouei, A., Breckon, T.P.: {G}anomaly: {S}emi-supervised anomaly detection via adversarial training. In: The Asian Conference on Computer Vision. pp. 622--637. Springer (2019)

\bibitem{chen2024enhancing}
Chen, S., Tang, Z., Cabezas, M., et~al.: Enhancing angular resolution via directionality encoding and geometric constraints in brain diffusion tensor imaging. ArXiv Preprint ArXiv:2409.07186  (2024)

\bibitem{ettehadi2022automated}
Ettehadi, N., Kashyap, P., Zhang, X., et~al.: Automated multiclass artifact detection in diffusion mri volumes via 3d residual squeeze-and-excitation convolutional neural networks. Frontiers in Human Neuroscience  \textbf{16},  877326 (2022)

\bibitem{he2016deep}
He, K., Zhang, X., Ren, S., et~al.: Deep residual learning for image recognition. In: Proceedings of the IEEE Conference on Computer Vision and Pattern Recognition (CVPR). pp. 770--778 (2016)

\bibitem{kingma2013auto}
Kingma, D.P., Welling, M.: Auto-encoding variational bayes. ArXiv Preprint ArXiv:1312.6114  (2013)

\bibitem{mcrobbie2017mri}
McRobbie, D.W., Moore, E.A., Graves, M.J., et~al.: {MRI} from {P}icture to {P}roton. Cambridge university press (2017)

\bibitem{michelucci2022introduction}
Michelucci, U.: An introduction to autoencoders. ArXiv Preprint ArXiv:2201.03898  (2022)

\bibitem{narayan1997generalized}
Narayan, S.: The generalized sigmoid activation function: {C}ompetitive supervised learning. Information Sciences  \textbf{99}(1-2),  69--82 (1997)

\bibitem{ni2006effects}
Ni, H., Kavcic, V., Zhu, T., et~al.: Effects of number of diffusion gradient directions on derived diffusion tensor imaging indices in human brain. American Journal of Neuroradiology  \textbf{27}(8),  1776--1781 (2006)

\bibitem{o2011introduction}
O’Donnell, L.J., Westin, C.F.: An introduction to diffusion tensor image analysis. Neurosurgery Clinics  \textbf{22}(2),  185--196 (2011)

\bibitem{ronneberger2015u}
Ronneberger, O., Fischer, P., Brox, T.: {U-Net}: {C}onvolutional networks for biomedical image segmentation. In: Proceedings of the International Conference on Medical Image Computing and Computer Assisted Intervention (MICCAI). pp. 234--241. Springer (2015)

\bibitem{samani2020qc}
Samani, Z.R., Alappatt, J.A., Parker, D., et~al.: {QC-Automator}: Deep learning-based automated quality control for diffusion {MR} images. Frontiers in Neuroscience  \textbf{13}, ~1456 (2020)

\bibitem{schlegl2019f}
Schlegl, T., Seeb{\"o}ck, P., Waldstein, S.M., et~al.: f-{A}no{GAN}: Fast unsupervised anomaly detection with generative adversarial networks. Medical Image Analysis  \textbf{54},  30--44 (2019)

\bibitem{tang2023high}
Tang, Z., Chen, S., D’Souza, A., et~al.: High angular diffusion tensor imaging estimation from minimal evenly distributed diffusion gradient directions. Frontiers in Radiology  \textbf{3},  1238566 (2023)

\bibitem{tang75reducing}
Tang, Z., Wang, X., Cabezas, M., et~al.: Reducing the impact of disrupted brain regions in diffusion tensor imaging with inpainting. In: Proceedings of the International Society of Magnetic Resonance in Medicine (ISMRM) (2023)

\bibitem{tang2022tw}
Tang, Z., Wang, X., Zhu, L., et~al.: Tw-bag: Tensor-wise brain-aware gate network for inpainting disrupted diffusion tensor imaging. In: Proceedings of the International Conference on Digital Image Computing: Techniques and Applications (DICTA). pp.~1--8. IEEE (2022)

\bibitem{tax2022s}
Tax, C.M., Bastiani, M., Veraart, J., et~al.: What’s new and what’s next in diffusion {MRI} preprocessing. NeuroImage  \textbf{249},  118830 (2022)

\bibitem{tournier2019diffusion}
Tournier, J.D.: Diffusion {MRI} in the brain--{T}heory and concepts. Progress in Nuclear Magnetic Resonance Spectroscopy  \textbf{112},  1--16 (2019)

\bibitem{tournier2019mrtrix3}
Tournier, J.D., Smith, R., Raffelt, D., et~al.: {MRtrix3}: {A} fast, flexible and open software framework for medical image processing and visualisation. NeuroImage  \textbf{202},  116137 (2019)

\bibitem{tournier2011diffusion}
Tournier, J.D., Mori, S., et~al.: Diffusion tensor imaging and beyond. Magnetic Resonance in Medicine  \textbf{65}(6), ~1532 (2011)

\bibitem{tsuruda1990diffusion}
Tsuruda, J.S., Chew, W.M., Moseley, M.E., et~al.: Diffusion-weighted {MR} imaging of the brain: value of differentiating between extraaxial cysts and epidermoid tumors. American Journal of Roentgenology  \textbf{155}(5),  1059--1065 (1990)

\bibitem{van2013wu}
Van~Essen, D.C., Smith, S.M., Barch, D.M., et~al.: The {WU}-{M}inn {H}uman {C}onnectome {P}roject: {A}n overview. NeuroImage  \textbf{80},  62--79 (2013)

\bibitem{victoroff1994method}
Victoroff, J., Mack, W., Grafton, S., et~al.: A method to improve interrater reliability of visual inspection of brain {MRI} scans in dementia. Neurology  \textbf{44}(12),  2267--2267 (1994)

\bibitem{wen2016discriminative}
Wen, Y., Zhang, K., Li, Z., et~al.: A discriminative feature learning approach for deep face recognition. In: Proceedings of the European Conference on Computer Vision (ECCV). pp. 499--515. Springer (2016)

\bibitem{zhao2021anomaly}
Zhao, H., Li, Y., He, N., et~al.: Anomaly detection for medical images using self-supervised and translation-consistent features. IEEE Transactions on Medical Imaging  \textbf{40}(12),  3641--3651 (2021)

\bibitem{zhu2017unpaired}
Zhu, J., Park, T., Isola, P., et~al.: Unpaired image-to-image translation using cycle-consistent adversarial networks. In: Proceedings of the IEEE International Conference on Computer Vision (ICCV). pp. 2223--2232 (2017)

\end{thebibliography}
\end{document}